\documentclass[aps,prc,reprint,superscriptaddress,nofootinbib]{revtex4-2}

\usepackage{graphicx}
\usepackage{amsmath}
\usepackage{bm}
\usepackage{float}
\usepackage{subcaption}

\usepackage[]{caption}
\usepackage{xcolor}
    \definecolor{myred}{HTML}{E45F62}
    \definecolor{myblue}{HTML}{5752D9}
    \definecolor{mygreen}{HTML}{318C45}

\usepackage{comment}
\renewcommand{\P}{\mathcal{P}}
\newcommand{\E}{\mathcal{E}}
\begin{document}


\title{Necessary and sufficient conditions of nonlinear causality \\ in viscous anisotropic hydrodynamics}


\author{Kento Yoshida}
\email{k-yoshida-4o6@eagle.sophia.ac.jp}
\affiliation{Department of Physics, Sophia University, Tokyo 102-8554, Japan}

\author{Shujun Zhao}
\email{zhaosj@sophia.ac.jp}
\affiliation{Department of Physics, Sophia University, Tokyo 102-8554, Japan}

\author{Tetsufumi Hirano}
\email{hirano@sophia.ac.jp}
\affiliation{Department of Physics, Sophia University, Tokyo 102-8554, Japan}



\begin{abstract}
We derive the necessary and sufficient conditions for nonlinear causality in viscous anisotropic hydrodynamics (VAH) within the approximation that neglects small correction terms.
Relativistic hydrodynamics provides a successful description of the space-time evolution of the matter produced in relativistic heavy-ion collisions, yet the earliest stage at which a hydrodynamic description becomes valid remains an open question. 
VAH has been proposed as an extension of conventional viscous hydrodynamics (VH) that can accommodate the large pressure anisotropies of the early-time dynamics.
However, in such far-from-equilibrium regimes, the nonlinear causality of the theory is not guaranteed.
By analyzing the characteristic velocities of the VAH equations of motion, we derive a set of inequalities that ensures causal signal propagation in all directions.
The resulting conditions take a remarkably simple form and admit a clear physical interpretation in terms of the characteristic modes of the anisotropic medium.
These results establish the regime of validity of VAH and provide a foundation for its application to the early-time dynamics of relativistic heavy-ion collisions.

\end{abstract}


\maketitle

\section{Introduction}
The main goal in the relativistic heavy-ion collisions at Relativistic Heavy Ion Collider (RHIC) and the Large Hadron Collider (LHC) is to create and study a novel QCD matter, the quark-gluon plasma (QGP). The observation of strong collectivity of the created QGP, reflected by the anisotropic flow~\cite{BRAHMS:2004adc, PHOBOS:2004zne, STAR:2005gfr, PHENIX:2004vcz, Gyulassy:2004vg, Muller:2006ee, 796947, Jacobs:2007dw, ALICE:2022wpn}, jet quenching~\cite{PHENIX:2001hpc, STAR:2003pjh, CMS:2011iwn, ATLAS:2010isq, ALICE:2010yje}, and strangeness enhancement~\cite{ALICE:2016fzo, Rafelski:1982pu}, suggests the strong coupling feature of the QGP matter.
The precise description of the soft observables in the hydrodynamic framework indicates that the generated QGP matter behaves like a nearly perfect liquid, with very small yet finite specific shear and bulk viscosities~\cite{Heinz:2013th, Gale:2013da, Song:2017wtw, Song:2013gia,  Bernhard:2019bmu, JETSCAPE:2020shq}.
In recent years, collectivity signals have also been observed in light-ion collisions~\cite{ALICE:2025luc, ATLAS:2025nnt, CMS:2025tga} or even in small collision systems~\cite{CMS:2012qk, ATLAS:2014qaj, ALICE:2012eyl, ALICE:2013snk, CMS:2014und, Zhao:2017rgg, Weller:2017tsr, Bozek:2013uha, ALICE:2016fzo, ALICE:2024vzv} at the LHC and RHIC.
The theoretical descriptions of the flow observables~\cite{Zhao:2020wcd, YuanyuanWang:2023wbz, Zhao:2017rgg, Zhao:2020pty, Schenke:2019pmk, Zhao:2025jwf} and strangeness enhancement~\cite{ALICE:2016fzo, Kanakubo:2019ogh,Kanakubo:2021qcw,Ito:2026hec} highly suggest the formation of a QGP droplet in light-ion and small collision systems.

However, the applicability of hydrodynamics is questionable during early-stage expansion or when the system is close to the freeze-out temperature.
In the early stage of evolution, the system will undergo rapid longitudinal expansion, driving it far from equilibrium~\cite{Berges:2020fwq, Keegan:2015avk, Fukushima:2016xgg}.
In heavy-ion collisions, such far-from-equilibrium effects can be largely washed out due to the attractor behavior of hydrodyncamic evolution~\cite{Romatschke:2017vte, Giacalone:2019ldn, Jankowski:2023fdz, Rajagopal:2024lou, Chen:2025qao}, while it may leave a fingerprint in the small collision systems in which the system expands more rapidly and experiences shorter evolution until freeze-out~\cite{Grosse-Oetringhaus:2024bwr}. 
When the system is close to transition between the QGP and hadronic matter, the bulk viscosity increases significantly, which may also result in the far-from-equilibrium evolution of the QGP system~\cite{Paech:2006st, Kharzeev:2007wb, Karsch:2007jc, Torrieri:2007fb, Torrieri:2008ip, Monnai:2009ad, Rajagopal:2009yw, Noronha-Hostler:2013gga}.
Together with these insights, various approaches have been developed to incorporate viscous effects non-perturbatively \cite{Bazow:2013ifa, Bazow:2015cha, Tinti:2015xwa, Molnar:2016vvu, Chesler:2015bba, Heller:2015dha,  McNelis:2018jho, Kurkela:2018wud, McNelis:2021zji}.
Among these, viscous anisotropic hydrodynamics (VAH) provides a theoretically consistent way to treat the anisotropic expansion in the early stage, as well as the bulk pressure non-perturbatively~\cite{Bazow:2013ifa, Bazow:2015cha, Tinti:2015xwa, Molnar:2016vvu, McNelis:2018jho, McNelis:2021zji}, largely extending the applicability of hydrodynamics in light-ions and small collision systems.

As the VAH framework incorporates the viscous corrections non-perturbatively, its validity in the far-from-equilibrium region requires further discussion on its phenomenological behavior and theoretical consistency.
The success of the VAH framework in describing the flows in heavy-ion collisions~\cite{McNelis:2021zji, Alqahtani:2017tnq, Liyanage:2023nds}
suggests the consistency between VAH and traditional viscous hydrodynamics when the system is close to thermalization in the final stage.
In small collision systems, its validity is further scrutinized by the correct description of the flow data ~\cite{Zhao:2025jwf}, which cannot be accurately described within the traditional viscous hydrodynamic framework. 
Moreover, it suggests that the VAH framework can reproduce the exact Boltzmann solution more accurately compared to the traditional framework~\cite{Bazow:2013ifa, Strickland:2017kux}.
The analysis of VAH evolution~\cite{McNelis:2021zji} suggests that the evolution of longitudinal pressure $\P_L$ in the VAH framework will always be positive, whereas in  traditional viscous hydrodynamics, it may not be satisfied due to the insufficient treatment of viscous corrections.
The discussions on the attractor behavior of VAH suggest that the VAH can capture the non-equilibrium attractor substantially better than conventional hydrodynamics~\cite{Strickland:2017kux, Cruz-Camacho:2018phl, Jaiswal:2022mdk}. 

Besides the above phenomenological and theoretical discussions, the investigation of the causality of VAH\footnote{We note that nonlinear causality in viscous anisotropic hydrodynamics was previously discussed in Ref.~\cite{Bemfica:2023res} within a conformal first-order framework. In contrast, the present work establishes the necessary and sufficient conditions for nonlinear causality in second-order anisotropic hydrodynamics. Accordingly, the underlying equations differ from those considered in Ref.~\cite{Bemfica:2023res}.} is also crucial for understanding its validity in the early stages of relativistic heavy-ion collisions.
The discussions of causality have a long history in relativistic dissipative hydrodynamics: Acausality and instability are known to arise in the relativistic Navier--Stokes (NS) equation of the first-order theory.
The further development of the second-order hydrodynamics~\cite{Israel:1976tn, Israel:1979wp, Muller:1967zza, Baier:2007ix,Monnai:2010qp,Denicol:2012cn} introduces relaxation processes in response to thermodynamic forces, which lead to additional non-hydrodynamic gapped modes, resulting in hyperbolic partial differential equations. The hydrodynamic equations obtained in this way are shown to obey causality, at least within a linear regime, as long as the relaxation time is sufficiently large~\cite{Hiscock:1983zz, Olson:1990rzl}.
The discussions further extend from the linearized hydrodynamic equations to the full nonlinear hydrodynamic equations.
A relation between causality and stability of the solutions of nonlinear hydrodynamic equations was investigated \cite{Denicol:2008ha,Pu:2009fj,Floerchinger:2017cii}.
Later, concrete criteria are obtained from the full nonlinear hydrodynamic equations in the nonlinear regime to identify the causal region of their solutions~\cite{Bemfica:2020xym}. 
Since these criteria provide inequalities that include thermodynamic variables and dissipative currents, they can be used to (in-)validate the numerical solutions of hydrodynamic equations \cite{Plumberg:2021bme,Chiu:2021muk,daSilva:2022xwu,ExTrEMe:2023nhy,Domingues:2024pom, Hoshino:2024qun,Roy:2025nay}.

In this paper, we investigate the nonlinear causality of the VAH framework. The papers are organized as follows. In Sec.~\ref{sec:eom}, we introduce the VAH framework and discuss the assumptions we made. In Sec.~\ref{sec:causality}, we derive the necessary and sufficient conditions that should be satisfied to preserve causality within the VAH framework. 
In Sec.~\ref{sec:discussion}, we discuss the interpretation of the necessary and sufficient conditions obtained in Sec.~\ref{sec:causality}.
Section \ref{sec:conclusion} summarizes our main conclusions and possible future directions.

The Minkowski metric is chosen to be $g_{\mu\nu}=\mathrm{diag}(1,-1,-1,-1)$ throughout this paper.

\section{Equations of motion}\label{sec:eom}
In the standard viscous hydrodynamics (VH), the energy-momentum tensor is decomposed as
\begin{align}
    T^{\mu\nu}=\E u^\mu u^\nu-(\P_{\rm eq}+\Pi)\Delta^{\mu\nu}+\pi^{\mu\nu},
\end{align}
where $\E$ is the energy density, $u^\mu$ is the flow velocity, $\P_{\rm eq}$ is the equilibrium pressure determined by the equation of state $\P_{\rm eq}=\P_{\rm eq}(\E)$, $\Pi$ is the bulk pressure, and $\pi^{\mu\nu}$ is the shear stress tensor. We take the definition of the spatial projection tensor as $\Delta^{\mu\nu}=g^{\mu\nu}-u^\mu u^\nu$.

In the viscous anisotropic hydrodynamics (VAH) framework, the energy-momentum tensor is decomposed alternatively as
\begin{align}
    T^{\mu\nu}=\E u^\mu u^\nu+\P_Lz^\mu z^\nu-\P_\perp\Xi^{\mu\nu}\notag\\
    +W^{\mu}_{\perp z}z^{\nu}+W^\nu_{\perp z}z^\mu+\pi_\perp^{\mu\nu},
\end{align}
where $\P_{L}$ ($\P_{\perp}$) is the longitudinal (transverse) pressure anisotropy, $W^{\mu}_{\perp z}$ is the longitudinal-momentum diffusion current, and $\pi^{\mu\nu}_{\perp}$ is the transverse shear-stress tensor. The longitudinal and transverse projections are taken using $z^\mu$ and $\Xi^{\mu\nu}=g^{\mu\nu}-u^\mu u^\nu+z^\mu z^\nu$, respectively. The projection is constructed by imposing the following orthogonal conditions:
\begin{align*}
    &u^\mu u_\mu=1,\quad z^\mu z_\mu=-1\quad u^\mu z_\mu=0,\\ 
    &\Xi^{\mu\nu}u_\nu=\Xi^{\mu\nu}z_\nu=0.
\end{align*}
Then the dissipative quantities $\pi^{\mu\nu}_\perp$ and $W^\mu_{\perp z}$ satisfy the constraints
\begin{align*}
\pi^{\mu}_{\perp\mu}=u_\mu\pi^{\mu\nu}_\perp=z_\mu\pi^{\mu\nu}_\perp=0,\quad u_\mu W^\mu_{\perp z}=z_\mu W^{\mu}_{\perp z}=0.
\end{align*}
Without introducing additional assumptions yet, the decomposition is equivalent to the VH one by re-decomposing the viscous terms as
\begin{align}
    \pi^{\mu\nu}&=\pi_\perp^{\mu\nu}+2W_{\perp z}^{(\mu}z^{\nu)}+\frac13(\P_L-\P_\perp)(2z^\mu z^\nu+\Xi^{\mu\nu}),\\
    \Pi&=\bar{\mathcal{P}}-\P_{\rm eq},
\end{align}
where $\bar{\mathcal{P}} = (2\P_\perp+\P_L)/3$.

In the VAH framework, however, the main assumption is that the evolution of anisotropic pressures $\P_{\perp,\ L}$ is treated non-perturbatively, with transport coefficients determined by the microscopic input. On the other hand, the residual stress tensor $\pi_\perp^{\mu\nu}$ and $W^\mu_{\perp,z}$ can be included as a small correction. Specifically, the microscopic distribution function is decomposed as $f=f_{a}+\delta\tilde{f}$, where the non-equilibrium leading correction $f_a$ incorporates the relaxation of the pressure anisotropy, while the next-to-leading-order (NLO) correction $\delta\tilde{f}$ incorporates the other viscous components, namely $\pi^{\mu\nu}_\perp$ and $W^\mu_{\perp z}$. Historically, the first-order transport coefficients are obtained from the kinetic theory calculations, with specific distribution functions chosen. The relaxation time, instead, is taken to be related to the isotropic transport coefficients $\eta$ and $\zeta$. In our work, we do not take such a specific assumption, but treat the transport coefficients as general phenomenological parameters, constrained by the nonlinear causality of the VAH framework.

In this paper, we focus on how the relaxation of pressure anisotropies in VAH framework affects the nonlinear causality. Thus, we ignore the NLO correction $\delta\tilde{f}$ and only consider the viscous corrections contained in $f_a$. The resulting equation of motion read

\begin{gather}
\dot{\E}+(\E+\P_\perp)\theta_\perp+(\E+\P_L)z_\mu D_z u^\mu=0,
\label{EoM-1}\\
(\E+\P_L)z_\mu \dot{u}^\mu= -D_z \P_L + (\P_L -\P_\perp)  \tilde{\theta}_\perp,\label{EoM-2}\\
(\E+\P_\perp)\Xi^\alpha_\mu\dot{u}^\mu = \nabla^\alpha_\perp \P_\perp + (\P_L-\P_\perp)\Xi^\alpha_\mu D_z z^\mu,
\label{EoM-3}\\
\dot{\P}_L=-\frac{\bar{\P}-\P_{\mathrm{eq}}}{\tau_\Pi}-\frac{\P_L-\P_\perp}{3\tau_\pi/2}+\bar{\zeta}^L_zz_\mu D_z u^\mu+\bar{\zeta}^L_\perp \theta_\perp,\label{EoM-4}\\
\dot{\P}_\perp=-\frac{\bar{\P}-\P_{\mathrm{eq}}}{\tau_\Pi}+\frac{\P_L-\P_\perp}{3\tau_\pi}+\bar{\zeta}^\perp_zz_\mu D_z u^\mu+\bar{\zeta}^\perp_\perp \theta_\perp.\label{EoM-5}
\end{gather}
Here $\theta_\perp=\nabla_{\perp,\mu}u^\mu$ and $\tilde{\theta}_\perp=\nabla_{\perp,\mu}z^\mu$ are the transverse expansion rates, with $\nabla_{\perp,\mu}=\Xi_{\mu\nu}\partial^\nu$. The derivative terms are defined as $\dot{A} = u^\mu\partial_\mu A$ and $D_z=-z^\mu\partial_\mu$. The expansion process of the pressure anisotropy $\P_{L,\perp}$ is controlled by the first-order anisotropic transport coefficients: $\bar{\zeta}^L_z$ ($\bar{\zeta}^\perp_\perp$) controls the longitudinal (transverse) expansion of the longitudinal (transverse) pressure $\P_L$ ($\P_\perp$), while the off-diagonal terms $\bar{\zeta}^L_\perp$ and $\bar{\zeta}^\perp_z$ control the diffusion among $\P_L$ and $\P_\perp$, respectively. In this paper, we introduce the shear and bulk relaxation time $\tau_{\pi}$ and $\tau_{\Pi}$, respectively, to account for the relaxation of the two viscous components.

Among the equations mentioned above, Eqs.~\eqref{EoM-1}--\eqref{EoM-3} represent the redecomposition of the energy-momentum conservation $\partial_\mu T^{\mu\nu}=0$ into the energy conservation and the momentum conservation along the longitudinal and transverse directions. Equations \eqref{EoM-4} and \eqref{EoM-5} represent the relaxation of the longitudinal and transverse pressure, respectively, toward the equilibrium limit.

These equations of motion contain only first-order derivatives with respect to space and time. Thus, by using a matrix representation, it can be written as a first-order quasilinear differential equation. We denote this by
\begin{equation}
    \mathbf{A}^\alpha (\mathbf{\Psi})\partial_\alpha\mathbf{\Psi}=\mathbf{F}(\mathbf{\Psi}).\label{matrix_representation}
\end{equation}
Here $\mathbf{A}^\alpha$ is an 8-by-8 coefficient matrix\footnote{It should be noted that each component of this matrix is Lorentz vector as shown in Appendix \ref{sec:matrix_represenation}.} and $\mathbf{\Psi}$ is an 8-dimensional column vector of unknown variables. The function $\mathbf{F}$ on the right-hand side represents terms including unknown variables $\mathbf{\Psi}$ without any space-time derivatives. 
The explicit matrix representation of Eq.~\eqref{matrix_representation} can be found in Appendix \ref{sec:matrix_represenation}.

\section{Causality}
\label{sec:causality}

Causality is a requirement that the propagation speed of information does not exceed the speed of light, and it is a fundamental principle of relativity theory. In second-order relativistic hydrodynamics under a sufficiently large relaxation time, the theory is constructed in such a way that the propagation of any information satisfies causality when the nonlinear equations of motion are linearized. However, it is not evident for the theory to satisfy causality at the nonlinear regime~\cite{Bemfica:2020xym}. Here, we derive the conditions under which the equations of motion of VAH \eqref{EoM-1}--\eqref{EoM-5} satisfy nonlinear causality.

First, in order to define the propagation speed, the equations of motion have to admit nontrivial solutions. The condition for this is
\begin{equation}
\det(\mathbf{A}^\alpha \xi_\alpha)=0, \label{definition_CE}
\end{equation}
using $\mathbf{A}^\alpha$ defined in Eq.~\eqref{matrix_representation}.
Equation \eqref{definition_CE} is called the \textit{characteristic equation}. The Lorentz vector $\xi_\alpha$ is a normal vector to the characteristic surfaces. Using a unit time-like vector $u_\alpha$, we decompose $\xi_\alpha$ as
\begin{equation}
    \xi_\alpha=bu_\alpha+a_\alpha,\label{xi_decompose}
\end{equation}
where $b(=u^\alpha \xi_\alpha)$ is a scalar, and $a_\alpha$ is a space-like vector satisfying $u^\alpha a_\alpha=0$. Clearly, in the local rest frame, it becomes $\xi_\alpha=(b,a_1,a_2,a_3)$.
From the gradient of characteristic surface in the local rest frame, the propagation speeds of the solutions can be defined as
\begin{equation}
    v_c^2=-\frac{b^2}{a^\alpha a_\alpha} = \frac{b^2}{a_1^2 +a_2^2 + a_3^3}.\label{definition_vc}
\end{equation}
This is called the \textit{characteristic velocity}.

Substituting the coefficient matrix of VAH shown in Eq.~\eqref{matrix} in Appendix \ref{sec:matrix_represenation} into the characteristic equation \eqref{definition_CE}, and taking the local rest frame for simplicity, we obtain
\begin{widetext}
\begin{equation}
    b^4+[\Gamma^\perp_\perp(a_1^2+a_2^2)+\Gamma^L_za_3^2]b^2+(\Gamma^L_z\Gamma^\perp_\perp-\Gamma^L_\perp\Gamma^\perp_z)(a_1^2+a_2^2)a_3^2=0,\label{CE_result1}
\end{equation}
where we have defined the following notations:
\begin{gather}
    \Gamma^L_z=\frac{\bar{\zeta}^L_z}{\E+\P_L},\
    \Gamma^\perp_\perp=\frac{\bar{\zeta}^\perp_\perp}{\E+\P_\perp},\
    \Gamma^L_\perp=\frac{\bar{\zeta}^L_\perp}{\E+\P_\perp},\
    \Gamma^\perp_z=\frac{\bar{\zeta}^\perp_z}{\E+\P_L}.
\end{gather}
Dividing both sides of equation \eqref{CE_result1} by $(a_1^2+a_2^2+a_3^2)^2$,
\begin{equation}
    v_c^4+[\Gamma^\perp_\perp(\hat{a}_1^2+\hat{a}_2^2)+\Gamma^L_z\hat{a}_3^2]v_c^2+(\Gamma^L_z\Gamma^\perp_\perp-\Gamma^L_\perp\Gamma^\perp_z)(\hat{a}_1^2+\hat{a}_2^2)\hat{a}_3^2=0.\label{CE_result2}
\end{equation}
\end{widetext}
Here $\hat{a}_i = a_i/\sqrt{a_1^2 + a_2^2 + a_3^2}$ is the $i$th component of the unit vector indicating its direction.
To satisfy causality is equivalent to requiring that, all solutions $v_c$ of Eq.~\eqref{CE_result2} satisfy $0\le|v_c|\le1$ for any $\hat{a}_i$. Since it contains only even-order terms of $v_c$, Eq.~\eqref{CE_result2} is a quadratic function of $v_c^2$.
Therefore, the necessary and sufficient condition for causality can be stated as follows:
\begin{center}
    \textit{For any $\hat{a}_i$, Eq.~\eqref{CE_result2} has two solutions in $0\le v_c^2\le1$.}
\end{center}
Let $f(v_c^2)$ denote the left-hand side of Eq.~\eqref{CE_result2}. Taking into account that the coefficient of the highest degree of the polynomial in $f(v_c^2)$ is positive and, consequently, that $f(v_c^2)$ is downwardly convex, the aforementioned condition can be understood graphically\footnote{While Eq.~\eqref{CE_result2} is a quadratic equation and its solutions can be obtained straightforwardly, deriving the necessary and sufficient conditions in this way is somewhat tedious.
Therefore, for completeness, the derivation following this approach is provided in Appendix~\ref{appendix:derivation2}.} as shown in Fig.~\ref{fig:quadraticfunction}.
Then, the condition above is equivalent to requiring that $f(v_c^2)$ satisfies the following four conditions for any $\hat{a}_i$:
\begin{itemize}
  \item[(a)] $f(v_c^2=0)\ge0$,
  \item[(b)] $f(v_c^2=1)\ge0$,
  \item[(c)] The symmetry axis of $f(v_c^2)$ lies in $0\le v_c^2\le1$,
  \item[(d)] The discriminant of $f(v_c^2)$ is $D\ge0$.
\end{itemize}
We derive the conditions that satisfy these four requirements one by one.

 \begin{figure}
 \includegraphics[width=0.4\textwidth]{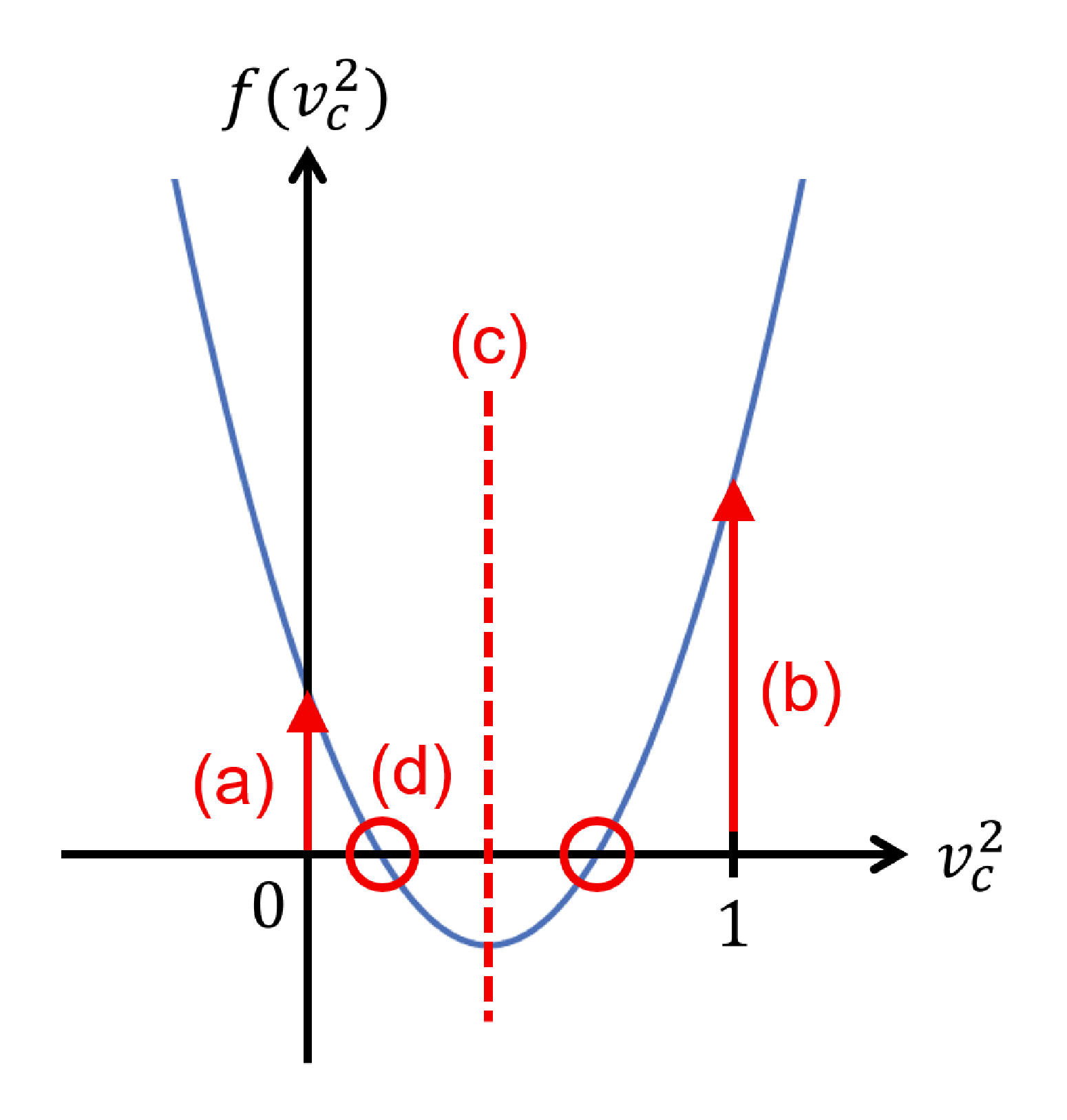}
 \caption{Conditions under which the quadratic function $f(v_c^2)=0$ in Eq.~\eqref{CE_result2} has two solutions in the range $0 \le v_c^2 \le 1$. The function $f(v_c^2)$ is shown by the blue curve, while the individual conditions (a)--(d) are indicated in red. \label{fig:quadraticfunction}}
 \end{figure}

Writing out these four conditions explicitly, we obtain as follows:
\begin{gather}
    (\Gamma^L_z\Gamma^\perp_\perp-\Gamma^L_\perp\Gamma^\perp_z)(\hat{a}_1^2+\hat{a}_2^2)\hat{a}_3^2\ge0,\label{condition_a}\\
    1+\Gamma^\perp_\perp(\hat{a}_1^2+\hat{a}_2^2) \nonumber \\
    +\Gamma^L_z\hat{a}_3^2+(\Gamma^L_z\Gamma^\perp_\perp-\Gamma^L_\perp\Gamma^\perp_z)(\hat{a}_1^2+\hat{a}_2^2)\hat{a}_3^2\ge0,\label{condition_b}\\
    0\le-\frac{1}{2}[\Gamma^\perp_\perp(\hat{a}_1^2+\hat{a}_2^2)+\Gamma^L_z\hat{a}_3^2]\le1,\label{condition_c}\\
     [\Gamma^\perp_\perp(\hat{a}_1^2+\hat{a}_2^2)+\Gamma^L_z\hat{a}_3^2]^2 \nonumber \\
     -4(\Gamma^L_z\Gamma^\perp_\perp-\Gamma^L_\perp\Gamma^\perp_z)(\hat{a}_1^2+\hat{a}_2^2)\hat{a}_3^2\ge0.\label{condition_d}
\end{gather}
Equation~\eqref{condition_a} can be solved straightforwardly.
By contrast, Eq.~\eqref{condition_b} itself takes the form of a quadratic inequality in $\hat{a}_i$, making it difficult to determine the condition under which it is satisfied for arbitrary $\hat{a}_i$.
However, by restricting the range to that obtained from Eq.~\eqref{condition_a}, the inequality becomes tractable and its solution can be determined.
Applying the restrictions obtained at each step successively to the subsequent conditions enables us to derive the necessary and sufficient conditions satisfying all four inequalities simultaneously.

For a detailed derivation, see Appendix \ref{appendix:derivation}. 
The necessary and sufficient conditions of nonlinear causality for equations of motion (\ref{EoM-1})--(\ref{EoM-5}) in viscous anisotropic hydrodynamics are finally summarized as follows:
\begin{subequations}\label{eq:gather}
\begin{gather}
-1\le\frac{\bar{\zeta}^L_z}{\E+\P_L}\le0,\label{eq:gathera}\\
-1\le\frac{\bar{\zeta}^\perp_\perp}{\E+\P_\perp}\le0,\label{eq:gatherb}\\
0\le\frac{\bar{\zeta}^L_\perp}{\E+\P_\perp}\frac{\bar{\zeta}^\perp_z}{\E+\P_L}\le\frac{\bar{\zeta}^L_z}{\E+\P_L}\frac{\bar{\zeta}^\perp_\perp}{\E+\P_\perp}\le1.\label{eq:gatherc}
\end{gather}
\end{subequations}
These are the main results of the present study.

\section{Discussion}\label{sec:discussion}
To clarify the physics behind Eq.~\eqref{eq:gather}, we parametrize the anisotropy by writing 
$a_\perp^2=a^2(1-\chi)$, $a_L^2=a^2\chi$, and $v_c^2\equiv b^2/a^2$, with $a^2 = a_1^2 + a_2^2 + a_3^2$ and $\chi\in[0,1]$. 
Equation~\eqref{CE_result2} then becomes
\begin{align}
v_c^4+&\bigl[\Gamma^\perp_\perp(1-\chi)+\Gamma^L_z\chi\bigr]v_c^2
\notag\\
\quad\quad+&\bigl(\Gamma^L_z\Gamma^\perp_\perp-\Gamma^L_\perp\Gamma^\perp_z\bigr)\chi(1-\chi)=0 .
\end{align}
Here $\chi$ measures the share of the longitudinal component in the propagation direction, with $\chi=0$ and $1$ corresponding to purely transverse and longitudinal propagation. The coefficients $\Gamma^L_\perp$ and $\Gamma^\perp_z$ describe the coupling (diffusion mixing) of the transverse and longitudinal propagations.

When the coupling vanishes, namely $\Gamma^L_\perp\Gamma^\perp_z=0$, the quartic factorizes as
\begin{align}
\bigl[v_c^2+\Gamma^\perp_\perp(1-\chi)\bigr]\bigl[v_c^2+\Gamma^L_z\chi\bigr]=0,
\end{align}
yielding two independent characteristic velocities $v_\perp^2=-\Gamma^\perp_\perp(1-\chi)$ and $v_L^2=-\Gamma^L_z\chi$. 
This factorization reveals an effective two-fluid structure: the system contains two independently propagating sound modes with distinct speeds, driven purely by the breaking of isotropy. Real, causal propagation requires $\Gamma^\perp_\perp,\Gamma^L_z\le0$, and imposing $0\le v_{\perp,L}^2\le1$ for all $\chi$ yields Eqs.~\eqref{eq:gathera} and \eqref{eq:gatherb}; their product then gives the leftmost inequality of Eq.~\eqref{eq:gatherc}.

Once $\Gamma^L_\perp\Gamma^\perp_z\neq0$, the coupling acts like a diffusion term between two fluids. A purely transverse or purely longitudinal initial disturbance no longer remains during propagation. Instead, the two sectors exchange pressure fluctuations continuously. The true eigenmodes of the system become the mixtures of the two components, determined by
\begin{align}
v_c^2=\frac12\Bigl(-\alpha\pm\sqrt{\alpha^2-4\beta}\Bigr),
\end{align}
where we define $\alpha=\Gamma^\perp_\perp(1-\chi)+\Gamma^L_z\chi$, $\beta=(\Gamma^L_z\Gamma^\perp_\perp-\Gamma^L_\perp\Gamma^\perp_z)\chi(1-\chi)$. The remaining inequalities in Eqs.~\eqref{eq:gatherc} acquire a clean physical meaning in this coupled picture. The fact that $v_c^2$ is the real number requires $\alpha^2-4\beta\ge0$, which is equivalent to
\begin{align}
    \Gamma^L_z\Gamma^\perp_\perp-\Gamma^L_\perp\Gamma^\perp_z\le\frac14\min_{\chi\in(0,1)}\left\{\frac{\alpha^2}{\chi(1-\chi)}\right\}=\Gamma^\perp_\perp\Gamma^L_z,
\end{align}
this implies $\Gamma^L_\perp\Gamma^\perp_z\ge0$. Positivity of the squared speeds further requires $\beta\ge0$, i.e.
\begin{align}
\Gamma^L_\perp\Gamma^\perp_z \le \Gamma^L_z\Gamma^\perp_\perp.
\end{align}
Combining these gives the chain Eq.~\eqref{eq:gatherc}:
\begin{align}
0\le \Gamma^L_\perp\Gamma^\perp_z \le \Gamma^\perp_\perp\Gamma^L_z \le 1 .
\end{align}

At the endpoint $\chi=0$ or $1$, the coupling term $\beta$ vanishes identically. Thus the velocities always reduce to the pure boundary values $v_\perp^2=-\Gamma^\perp_\perp$ and $v_L^2=-\Gamma^L_z$, irrespective of $\Gamma^L_\perp\Gamma^\perp_z$. The mixing coefficients thus govern only the interior $\chi\in(0,1)$. If $\Gamma^L_\perp\Gamma^\perp_z=0$, the pure modes persist as exact eigenmodes for all $\chi$, recovering the decoupled two-fluid picture discussed above. If the coupling is active yet satisfies Eq.~\eqref{eq:gatherc}, the two branches interpolate smoothly between their fixed endpoint values, remaining real and non-negative throughout. Consequently, Eqs.~\eqref{eq:gathera} and \eqref{eq:gatherb} enforce causality at the limiting propagation directions, while Eq.~\eqref{eq:gatherc} guarantees that these constraints extend consistently to the fully mixed velocity $v_c^2(\chi)$ for every intermediate angle.

\section{Conclusions}\label{sec:conclusion}

In this paper, we derived the necessary and sufficient conditions for nonlinear causality under which the equations of motion in the viscous anisotropic hydrodynamics (VAH). These conditions guarantee that the VAH equations of motion obey the relativistic causality in the nonlinear regime. 
Consequently, the applicability of VAH is limited to regions of parameter space where these conditions are satisfied.
While the necessary and sufficient conditions previously obtained for viscous hydrodynamics (VH) are highly involved, the corresponding conditions derived here for VAH take a remarkably simple form. This simplification may be attributed to the neglect of next-to-leading-order (NLO) dissipative corrections.

Our analysis of the characteristic velocities reveals that the breaking of isotropic symmetry splits the system into two distinct eigenmodes.
The resulting constraints ensure that, for all propagation directions, the mixed modes possess real and non-negative squared characteristic velocities. 
As a consequence, both thermodynamic stability and causality are maintained throughout the anisotropic medium.

A natural next step is to clarify the relationship between VAH and conventional VH in order to better understand the physical implications of the causality conditions derived here.
It is also important to incorporate the NLO correction, $\delta\tilde{f}$, into the equations of motion and establish the necessary and sufficient conditions for nonlinear causality in the complete VAH framework.
Ultimately, we plan to perform simulations of high-energy heavy-ion collisions within VAH and compare the results with those obtained from VH. Such a study will provide valuable insight into the applicability of hydrodynamic descriptions during the early stages of the collision evolution.

\begin{acknowledgements}
We gratefully acknowledge Y.~Nara for suggesting this line of investigation.
The work by T.H. was supported by JSPS KAKENHI Grant No.~JP23K03395.    
\end{acknowledgements}

\appendix

\section{Matrix Representation}\label{sec:matrix_represenation}

We explicitly show the matrix representation of the equations of motion of VAH \eqref{EoM-1}--\eqref{EoM-5}. These can be compiled in matrix form as follows:

\begin{widetext}

\begin{equation}
\begin{split}
\begin{pmatrix}
u^\alpha & 0 & 0 & A^\alpha_0-Bu^zz^\alpha & A^\alpha_1 & A^\alpha_2 & A^\alpha_3+Bu^tz^\alpha & 0 \\
0 & z^\alpha & 0 & C^\alpha_3-Bu^zu^\alpha & 0 & 0 & C^\alpha_0+Bu^tu^\alpha & D^\alpha \\
0 & 0 & -\Xi^{\alpha0} & A^0_0u^\alpha+C^0_3z^\alpha & A^0_1u^\alpha & A^0_2u^\alpha & A^0_3u^\alpha+C^0_0z^\alpha & D^0z^\alpha \\
0 & 0 & -\Xi^{\alpha1} & A^1_0u^\alpha+C^1_3z^\alpha & A^1_1u^\alpha & A^1_2u^\alpha & A^1_3u^\alpha+C^1_0z^\alpha & D^1z^\alpha\\
0 & 0 & -\Xi^{\alpha2} & A^2_0u^\alpha+C^2_3z^\alpha & A^2_1u^\alpha & A^2_2u^\alpha & A^2_3u^\alpha+C^2_0z^\alpha & D^2z^\alpha\\
0 & 0 & -\Xi^{\alpha3} & A^3_0u^\alpha+C^3_3z^\alpha & A^3_1u^\alpha & A^3_2u^\alpha & A^3_3u^\alpha+C^3_0z^\alpha & D^3z^\alpha\\
0 & -u^\alpha & 0 & \bar{\zeta}^L_\perp\Xi^\alpha_0-\bar{\zeta}^L_z\gamma_L u^zz^\alpha & \bar{\zeta}^L_\perp\Xi^\alpha_1 & \bar{\zeta}^L_\perp\Xi^\alpha_2 & \bar{\zeta}^L_\perp\Xi^\alpha_3+\bar{\zeta}^L_z\gamma_L u^tz^\alpha & 0\\
0 & 0 & -u^\alpha & \bar{\zeta}^\perp_\perp\Xi^\alpha_0-\bar{\zeta}^\perp_z\gamma_L u^zz^\alpha & \bar{\zeta}^\perp_\perp\Xi^\alpha_1 & \bar{\zeta}^\perp_\perp\Xi^\alpha_2 & \bar{\zeta}^\perp_\perp\Xi^\alpha_3+\bar{\zeta}^\perp_z\gamma_L u^tz^\alpha & 0
\end{pmatrix}
\partial_\alpha
\begin{pmatrix}
\E\\
\P_L\\
\P_\perp\\
u^t\\
u^x\\
u^y\\
u^z\\
\gamma_L
\end{pmatrix}
=
\begin{pmatrix}
0\\
0\\
0\\
0\\
0\\
0\\
\frac{\bar{\P}-\P_{\mathrm{eq}}}{\tau_\Pi}+\frac{\P_L-\P_\perp}{3\tau_\pi/2}\\
\frac{\bar{\P}-\P_{\mathrm{eq}}}{\tau_\Pi}-\frac{\P_L-\P_\perp}{3\tau_\pi}
\end{pmatrix}.
\end{split}\label{matrix}
\end{equation}

\end{widetext}

Here, we define the following notations:
\begin{gather}
A^\mu_\nu=(\E+\P_\perp)\Xi^\mu_\nu,\\
B=(\E+\P_L)\gamma_L,\\
C^\mu_\nu=(\P_L-\P_\perp)\Xi^\mu_\nu\gamma_L,\\
D^\mu=(\P_L-\P_\perp)(\Xi^\mu_0u^z+\Xi^\mu_3u^t),
\end{gather}
where we employ a parametrization of longitudinal projection as $z^\mu = \gamma_L(u^z, 0, 0, u^t)$ and $\gamma_L = 1/\sqrt{(u^t)^2 - (u^z)^2}$.
The coefficient matrix in Eq.~\eqref{matrix} is $8\times8$ matrix with respect to Lorentz vector. Multiplying this matrix by $\xi_\alpha$ in Eq.~\eqref{xi_decompose}, we obtain $\mathbf{A}^\alpha\xi_\alpha$. In addition, we take the local rest frame in which $u^\alpha = (1, 0, 0, 0)$, $z^\alpha = (0, 0, 0, 1)$, and $\xi_\alpha=(b,a_1,a_2,a_3)$. 
As a result, the number of independent variables and equations is reduced by two, thus the matrix eventually becomes a $6\times6$ one. Substituting the resulting $\mathbf{A}^\alpha\xi_\alpha$ into Eq.~\eqref{definition_CE}, we obtain the characteristic equation as:

\begin{widetext}
\begin{equation}
\det(\mathbf{A}^\alpha\xi_\alpha)=\det
\begin{pmatrix}
b & 0 & 0 & (\E+\P_\perp)a_1 & (\E+\P_\perp)a_2 & (\E+\P_L)a_3\\
0 & a_3 & 0 & 0 & 0 & (\E+\P_L)b\\
0 & 0 & a_1 & (\E+\P_\perp)b & 0 & 0\\
0 & 0 & a_2 & 0 & (\E+\P_\perp)b & 0\\
0 & -b & 0 & \bar{\zeta}^L_\perp a_1 & \bar{\zeta}^L_\perp a_2 & \bar{\zeta}^L_z a_3\\
0 & 0 & -b & \bar{\zeta}^\perp_\perp a_1 & \bar{\zeta}^\perp_\perp a_2 & \bar{\zeta}^\perp_z a_3
\end{pmatrix}
=0.\label{result_CE1}
\end{equation}

\end{widetext}

By carrying out this calculation, we obtain Eq.~\eqref{CE_result1}. Note that the trivial solution $b=0$ has been excluded in the course of the calculation.

\section{Derivation of Conditions}\label{appendix:derivation}

We present the procedure for deriving the necessary and sufficient conditions from the characteristic equation \eqref{result_CE1}.
Let $f(v_c^2)$ denote the left-hand side of equation \eqref{CE_result2}; that is, define

\begin{widetext}
\begin{equation}
  f(v_c^2)=  v_c^4+[\Gamma^\perp_\perp(\hat{a}_1^2+\hat{a}_2^2)+\Gamma^L_z\hat{a}_3^2]v_c^2+(\Gamma^L_z\Gamma^\perp_\perp-\Gamma^L_\perp\Gamma^\perp_z)(\hat{a}_1^2+\hat{a}_2^2)\hat{a}_3^2.\label{f_v_c_2}
\end{equation}

\end{widetext}
The necessary and sufficient condition for causality is equivalent to requiring that $f(v_c^2)$ satisfy the following four conditions for any $\hat{a}_i$:
\begin{itemize}
  \item[(a)] $f(v_c^2=0)\ge0$,
  \item[(b)] $f(v_c^2=1)\ge0$,
  \item[(c)] The symmetry axis of $f(v_c^2)$ lies in $0\le v_c^2\le1$,
  \item[(d)] The discriminant of $f(v_c^2)$ is $D\ge0$.
\end{itemize}
We derive the conditions that satisfy these four requirements one by one.

\subsection{Preparation}

Since $\hat{a}_i (= a_i/\sqrt{a_1^2 + a_2^2 + a_3^2})$ is a component of a unit vector, we have
\begin{equation}
    \hat{a}_1^2+\hat{a}_2^2=1-\hat{a}_3^2.
\end{equation}
Each component $\hat{a}_i$ taking any value is equivalent to the square of the third component $\hat{a}_3^2$ taking any value in $0\le\hat{a}_3^2\le1$.
Substituting this into Eq.~\eqref{f_v_c_2}, $f(v_c^2)$ can be written as follows:
\begin{widetext}
\begin{equation}
    f(v_c^2)=(v_c^2)^2+[\Gamma^\perp_\perp(1-\hat{a}_3^2)+\Gamma^L_z\hat{a}_3^2]v_c^2+(\Gamma^L_z\Gamma^\perp_\perp-\Gamma^L_\perp\Gamma^\perp_z)(1-\hat{a}_3^2)\hat{a}_3^2.\label{CE_result3}
\end{equation}
\end{widetext}
Thus, $f(v_c^2)$ depends only on $v_c^2$ and $\hat{a}_3^2$. Therefore, it is necessary and sufficient for $f(v_c^2)$ to satisfy the above conditions (a)--(d) for any $\hat{a}_3^2$ such that $0\le\hat{a}_3^2\le1$.

\subsection{Condition (a)}

First, we derive the condition under which Condition (a), namely $f(v_c^2=0)\ge0$, is satisfied.
Substituting $v_c^2=0$ into $f(v_c^2)$, we obtain
\begin{equation}
    f(0)=(\Gamma^L_z\Gamma^\perp_\perp-\Gamma^L_\perp\Gamma^\perp_z)(1-\hat{a}_3^2)\hat{a}_3^2.
\end{equation}
Since $(1-\hat{a}_3^2)\hat{a}_3^2$ is positive, for $f(v_c^2=0)\ge0$ to hold for any $\hat{a}_3^2$, it suffices that
\begin{equation}
    \Gamma^L_z\Gamma^\perp_\perp-\Gamma^L_\perp\Gamma^\perp_z\ge0.
\end{equation}
Therefore, the condition for satisfying (a) is as follows:
\begin{equation}
    \Gamma^L_\perp\Gamma^\perp_z\le\Gamma^L_z\Gamma^\perp_\perp.
\end{equation}

\subsection{Condition (a)+(b)}
Next, we derive the condition under which Condition (b), namely $f(v_c^2=1)\ge0$, is satisfied together with Condition (a) obtained in the first step mentioned above.

Substituting $v_c^2=1$ into $f(v_c^2)$ and rearranging in terms of $\hat{a}_3^2$, we obtain
\begin{widetext}
\begin{equation}
f(1)=-(\Gamma^L_z\Gamma^\perp_\perp-\Gamma^L_\perp\Gamma^\perp_z)(\hat{a}_3^2)^2+(\Gamma^L_z-\Gamma^\perp_\perp+\Gamma^L_z\Gamma^\perp_\perp-\Gamma^L_\perp\Gamma^\perp_z)\hat{a}_3^2+\Gamma^\perp_\perp+1. \label{g_of_b}
\end{equation}
\end{widetext}
Now, let us define the right-hand side of Eq.~\eqref{g_of_b} be the quadratic function as $g(\hat{a}_3^2)$. From Condition (a) obtained above, we have $\Gamma^L_z\Gamma^\perp_\perp-\Gamma^L_\perp\Gamma^\perp_z\ge0$. Therefore, the leading coefficient of $g(\hat{a}_3^2)$ is negative. Then, in order to satisfy $f(v_c^2=1)\ge0$, it suffices that $g(\hat{a}_3^2)$ satisfy \textit{both} of the following two conditions:
\begin{description}
  \item[(i)] $g(\hat{a}_3^2=0)\ge0$.
  
  Substituting $\hat{a}_3^2=0$ into $g(\hat{a}_3^2)$, we obtain
  \begin{equation}
      g(0)=\Gamma^\perp_\perp+1.
  \end{equation}
  Then, the condition for $g(\hat{a}_3^2=0)\ge0$ is given by
  \begin{equation}
      -1\le\Gamma^\perp_\perp.
  \end{equation}
  
  \item[(ii)] $g(\hat{a}_3^2=1)\ge0$.

  Substituting $\hat{a}_3^2=1$ into $g(\hat{a}_3^2)$, we obtain
  \begin{equation}
      g(1)=\Gamma^L_z+1.
  \end{equation}
  Then, the condition for $g(\hat{a}_3^2=1)\ge0$ is given by
  \begin{equation}
      -1\le\Gamma^L_z.
  \end{equation} 
  
\end{description}

Therefore, the conditions for satisfying Conditions (a) and (b) are as follows:
\begin{gather*}
     -1\le\Gamma^L_z,\\
     -1\le\Gamma^\perp_\perp,\\
     \Gamma^L_\perp\Gamma^\perp_z\le\Gamma^L_z\Gamma^\perp_\perp.
 \end{gather*}

\subsection{Condition (a)+(b)+(c)}

Next, we derive the condition under which Condition (c), namely the symmetry axis of $f(v_c^2)$ lies in $0\le v_c^2\le1$, is satisfied together with Conditions (a) and (b).

The symmetry axis of $f(v_c^2)$ is given by
\begin{equation}
\begin{split}
v_c^2&=-\frac{1}{2}[\Gamma^\perp_\perp(1-\hat{a}_3^2)+\Gamma^L_z\hat{a}_3^2]\\
&=\frac{1}{2}[(\Gamma^\perp_\perp-\Gamma^L_z)\hat{a}_3^2-\Gamma^\perp_\perp].
\end{split}
\end{equation}
Then, the condition for $0\le v_c^2\le1$ is
\begin{equation}
0\le(\Gamma^\perp_\perp-\Gamma^L_z)\hat{a}_3^2-\Gamma^\perp_\perp\le2.
\end{equation}
Now, let the intermediate expression between two inequalities be the linear function $h(\hat{a}_3^2)$. Then, the condition is divided into two cases according to the sign of the coefficient of $h(\hat{a}_3^2)$:

\begin{description}
  \item[(i)] In the case $\Gamma^L_z\le\Gamma^\perp_\perp$.
  
The linear function $h(\hat{a}_3^2)$ takes the minimum value $-\Gamma^\perp_\perp$ at $\hat{a}_3^2=0$, and the maximum value $-\Gamma^L_z$ at $\hat{a}_3^2=1$. Therefore, the conditions are the following two inequalities:
  \begin{gather}
0\le-\Gamma^\perp_\perp,\\
-\Gamma^L_z\le2.
  \end{gather}
  
  \item[(ii)] In the case $\Gamma^\perp_\perp<\Gamma^L_z$.

The linear function $h(\hat{a}_3^2)$ takes the minimum value $-\Gamma^L_z$ at $\hat{a}_3^2=1$, and the maximum value $-\Gamma^\perp_\perp$ at $\hat{a}_3^2=0$. Therefore, the conditions are the following two inequalities:
  \begin{gather}
0\le-\Gamma^L_z,\\
-\Gamma^\perp_\perp\le2.
  \end{gather}
  
\end{description}

These conditions obtained in the separate cases can be combined into the following conditions:
\begin{gather}
    -2\le\Gamma^L_z\le0,\\
    -2\le\Gamma^\perp_\perp\le0.
\end{gather}

Therefore, the conditions for satisfying (a)+(b)+(c) are as follows:
\begin{gather*}
     -1\le\Gamma^L_z\le0,\\
     -1\le\Gamma^\perp_\perp\le0,\\
     \Gamma^L_\perp\Gamma^\perp_z\le\Gamma^L_z\Gamma^\perp_\perp\le1.
 \end{gather*}

\subsection{Condition (a)+(b)+(c)+(d)}

Finally, we derive the condition under which Condition (d), namely the discriminant of $f(v_c^2)$ is $D\ge0$, is satisfied together with Conditions (a), (b) and (c).

The discriminant of $f(v_c^2)$ is given by
\begin{align}
D=[\Gamma^\perp_\perp(1-\hat{a}_3^2)+\Gamma^L_z\hat{a}_3^2]^2-4(\Gamma^L_z\Gamma^\perp_\perp-\Gamma^L_\perp\Gamma^\perp_z)(1-\hat{a}_3^2)\hat{a}_3^2.
\end{align}
Here, let $t=\hat{a}_3^2/1-\hat{a}_3^2$. Then, since $\hat{a}_3^2=t/1+t$ and $1-\hat{a}_3^2=1/1+t$, discriminant is given by
\begin{equation}
\begin{split}
D&=\frac{(\Gamma^\perp_\perp+\Gamma^L_zt)^2-4(\Gamma^L_z\Gamma^\perp_\perp-\Gamma^L_\perp\Gamma^\perp_z)t}{(1+t)^2}
\\
&=\frac{(\Gamma^L_z)^2t^2+(4\Gamma^L_\perp\Gamma^\perp_z-2\Gamma^L_z\Gamma^\perp_\perp)t+(\Gamma^\perp_\perp)^2}{(1+t)^2}.
\end{split}
\end{equation}

When $\hat{a}_3^2$ takes any value in $0\le\hat{a}_3^2\le1$, $t$ takes any value in $0\le t<+\infty$. That is, we derive the condition that $D\ge0$ for all $t\ge0$. Since the denominator is clearly positive, the sign of $D$ is same as that of the numerator. Let the numerator be the quadratic function $k(t)$. Since the leading coefficient of $k(t)$ is positive, in order to satisfy $k(t)\ge0$ for all $t\ge0$, it suffices that $k(t)$ satisfy \textit{either} of the following two conditions:

\begin{description}
  \item[(i)] The discriminant of $k(t)$ is $D_{(k)}\le0$.

  The discriminant of $k(t)$ is given by
  \begin{equation}
  \begin{split}
      D_{(k)}/4&=(2\Gamma^L_\perp\Gamma^\perp_z-\Gamma^L_z\Gamma^\perp_\perp)^2-(\Gamma^L_z\Gamma^\perp_\perp)^2\\
      &=4\Gamma^L_\perp\Gamma^\perp_z(\Gamma^L_\perp\Gamma^\perp_z-\Gamma^L_z\Gamma^\perp_\perp).
  \end{split}
  \end{equation}
  Here, from the condition (a) obtained above, we have $\Gamma^L_\perp\Gamma^\perp_z-\Gamma^L_z\Gamma^\perp_\perp\le0$. Then, the condition for $D_{(k)}\le0$ is given by
  \begin{equation}
      \Gamma^L_\perp\Gamma^\perp_z\ge0.
  \end{equation}
  
  \item[(ii)] $k(t=0)\ge0$, and the symmetry axis of $k(t)$ lies in $t\le0$.

  The former, $k(t=0)\ge0$ is obvious. The symmetry axis of $k(t)$ is given by
  \begin{equation}
      t=-\frac{2\Gamma^L_\perp\Gamma^\perp_z-\Gamma^L_z\Gamma^\perp_\perp}{(\Gamma^L_z)^2}
  \end{equation}
  Since the denominator is clearly positive, the condition for $t\le0$ is
  \begin{equation}
      2\Gamma^L_\perp\Gamma^\perp_z-\Gamma^L_z\Gamma^\perp_\perp\ge0
  \end{equation}
  
\end{description}

Under the condition (a), if $\Gamma^L_\perp\Gamma^\perp_z\le0$, then  $\Gamma^L_\perp\Gamma^\perp_z\le\Gamma^L_z\Gamma^\perp_\perp$ implies  $2\Gamma^L_\perp\Gamma^\perp_z-\Gamma^L_z\Gamma^\perp_\perp\le0$. Hence, in order to satisfy the condition (i) or (ii), it is necessary that $\Gamma^L_\perp\Gamma^\perp_z\ge0$.

Therefore, the conditions for satisfying all (a) to (d) are as follows:
\begin{gather*}
     -1\le\Gamma^L_z\le0,\\
     -1\le\Gamma^\perp_\perp\le0,\\
     0\le\Gamma^L_\perp\Gamma^\perp_z\le\Gamma^L_z\Gamma^\perp_\perp\le1.
 \end{gather*}

\subsection{Results}

Writing $\Gamma$ explicitly, we finally obtain the following results:
\begin{gather}
-1\le\frac{\bar{\zeta}^L_z}{\E+\P_L}\le0,\\
-1\le\frac{\bar{\zeta}^\perp_\perp}{\E+\P_\perp}\le0,\\
0\le\frac{\bar{\zeta}^L_\perp}{\E+\P_\perp}\frac{\bar{\zeta}^\perp_z}{\E+\P_L}\le\frac{\bar{\zeta}^L_z}{\E+\P_L}\frac{\bar{\zeta}^\perp_\perp}{\E+\P_\perp}\le1.
\end{gather}
These are the necessary and sufficient conditions for nonlinear causality in VAH.

\section{Another Approach to Deriving Conditions}\label{appendix:derivation2}

So far, we have adopted a geometric approach based on Fig.~\ref{fig:quadraticfunction} to derive the conditions. However, since Eq.~\eqref{CE_result2} is a quadratic equation of $v_c^2$, it is also possible to obtain the solutions directly and impose the causality conditions on them. Here, we present this approach.

First, for simplicity, we derive the conditions that a general quadratic equation $x^2+\alpha x+\beta=0$, whose leading coefficient is $1$, has real solutions, and all of them satisfy $0\le x\le1$.
Using the quadratic formula, the solutions for $x$ are given by
\begin{equation}
    x=\frac{1}{2}\Bigl(-\alpha\pm\sqrt{\alpha^2-4\beta}\Bigr)\label{solution}.
\end{equation}
First, the condition for these solutions to be real is
\begin{equation}
    \alpha^2-4\beta\ge0\label{discriminant}.
\end{equation}
This constitutes the first condition. Among the two solutions given by Eq.~\eqref{solution}, let $x_1$ denote the solution obtained by choosing the minus sign in the $\pm$ term, and $x_2$ denote the one obtained by choosing plus sign. Then, since it is clear that $x_1\le x_2$, it suffices to require $0\le x_1$ and $x_2\le1$ in order for all solutions to satisfy $0\le x\le1$. From $0\le x_1$, we obtain
\begin{gather}
    0\le -\alpha-\sqrt{\alpha^2-4\beta},\\
    \alpha\le-\sqrt{\alpha^2-4\beta}\label{alpha}.
\end{gather}
Since the right-hand side is clearly negative under the condition of Eq.~\eqref{discriminant}, we obtain
\begin{equation}
    \alpha\le0.
\end{equation}
This constitutes the second condition. Noting that both sides are negative, we further rearrange Eq.~\eqref{alpha} to obtain
\begin{gather}
    \alpha^2\ge \alpha^2-4\beta,\\
    0\ge-4\beta\label{beta}.
\end{gather}
Then, we obtain
\begin{equation}
    \beta\ge0
\end{equation}
This constitutes the third condition. From $x_2\le1$, we obtain
\begin{gather}
    2\ge -\alpha+\sqrt{\alpha^2-4\beta},\\
    \alpha+2\ge\sqrt{\alpha^2-4\beta}.
\end{gather}
Noting that both sides are positive,
\begin{gather}
    (\alpha+2)^2\ge\alpha^2-4\beta,\\
    4\alpha+4\beta+4\ge0.
\end{gather}
Then, we obtain
\begin{equation}
    \alpha+\beta\ge-1.
\end{equation}
This constitutes the fourth condition.

Thus, four conditions have been derived in the general form. From here, we substitute the explicit expressions into these. From Eq.~\eqref{CE_result1}, the explicit forms of $\alpha$ and $\beta$ are given by
\begin{gather}
    \alpha=\Gamma^\perp_\perp(\hat{a}_1^2+\hat{a}_2^2)+\Gamma^L_z\hat{a}_3^2,\\
    \beta=(\Gamma^L_z\Gamma^\perp_\perp-\Gamma^L_\perp\Gamma^\perp_z)(\hat{a}_1^2+\hat{a}_2^2)\hat{a}_3^2.
\end{gather}
Substituting these into the above conditions, we obtain the following constraints:
\begin{widetext}
\begin{gather}
     [\Gamma^\perp_\perp(\hat{a}_1^2+\hat{a}_2^2)+\Gamma^L_z\hat{a}_3^2]^2-4(\Gamma^L_z\Gamma^\perp_\perp-\Gamma^L_\perp\Gamma^\perp_z)(\hat{a}_1^2+\hat{a}_2^2)\hat{a}_3^2\ge0,\label{condition2_d}\\
      \Gamma^\perp_\perp(\hat{a}_1^2+\hat{a}_2^2)+\Gamma^L_z\hat{a}_3^2\le0,\label{condition2_c}\\
       (\Gamma^L_z\Gamma^\perp_\perp-\Gamma^L_\perp\Gamma^\perp_z)(\hat{a}_1^2+\hat{a}_2^2)\hat{a}_3^2\ge0,\label{condition2_a}\\
       \Gamma^\perp_\perp(\hat{a}_1^2+\hat{a}_2^2)+\Gamma^L_z\hat{a}_3^2+(\Gamma^L_z\Gamma^\perp_\perp-\Gamma^L_\perp\Gamma^\perp_z)(\hat{a}_1^2+\hat{a}_2^2)\hat{a}_3^2\ge-1.\label{condition2_b}
\end{gather}
\end{widetext}
Equations~\eqref{condition2_d},~\eqref{condition2_c},~\eqref{condition2_a}, and~\eqref{condition2_b} are identical to Eqs.~\eqref{condition_d},~\eqref{condition_c},~\eqref{condition_a}, and~\eqref{condition_b}, respectively.
Therefore, we obtain the same conditions as in the graphical approach. Then, the subsequent discussion is the same. Note that Eq.~\eqref{condition2_c} takes a slightly different from here, with only one bound appearing instead of both upper and lower bounds. Nevertheless, the final result is the same after calculation.

\bibliography{ref}
\end{document}